\documentclass[sigconf]{acmart}

\AtBeginDocument{%
  \providecommand\BibTeX{{%
    \normalfont B\kern-0.5em{\scshape i\kern-0.25em b}\kern-0.8em\TeX}}}

\copyrightyear{2021} 
\acmYear{2021} 
\setcopyright{acmcopyright}\acmConference[WI-IAT '21]{IEEE/WIC/ACM International Conference on Web Intelligence}{December 14--17, 2021}{ESSENDON, VIC, Australia}
\acmBooktitle{IEEE/WIC/ACM International Conference on Web Intelligence (WI-IAT '21), December 14--17, 2021, ESSENDON, VIC, Australia}
\acmPrice{15.00}
\acmDOI{10.1145/3486622.3493984}
\acmISBN{978-1-4503-9115-3/21/12}

\usepackage{graphicx}
\usepackage{diagbox}
\usepackage{multirow}

\acmSubmissionID{WI277}

\usepackage{color}

\begin{document}

\title[Online Classroom Evaluation System Based on Multi-Reaction Estimation]
{Online Classroom Evaluation System Based on Multi-Reaction Estimation}

\author{Yanyi Peng}
\affiliation{%
  \institution{Department of Computer Science, Graduate School of Engineering, Nagoya Institute of Technology}
  \streetaddress{Gokiso-cho, Showa-ku}
  \city{Nagoya}
  \state{Aichi}
  \country{Japan}
  \postcode{466-8555}
}
\email{pyy@toralab.org}

\author{Masato Kikuchi}
\affiliation{%
  \institution{Nagoya Institute of Technology}
  \streetaddress{Gokiso-cho, Showa-ku}
  \city{Nagoya}
  \state{Aichi}
  \country{Japan}
  \postcode{466-8555}
}
\email{kikuchi@nitech.ac.jp}

\author{Tadachika Ozono}
\affiliation{%
  \institution{Nagoya Institute of Technology}
  \streetaddress{Gokiso-cho, Showa-ku}
  \city{Nagoya}
  \state{Aichi}
  \country{Japan}
  \postcode{466-8555}
}
\email{ozono@nitech.ac.jp}

\renewcommand{\shortauthors}{Peng, et al.}

\begin{abstract}
  Online learning is more convenient than traditional face-to-face teaching methods. However, during real-time online classes, it is difficult for teachers to observe the reactions of all students simultaneously. Herein, we introduce an online education classroom evaluation system that enables teachers to adjust the speed of their lessons based on students’ reactions. We aim to develop a method that can evaluate student participation based on multi-reaction of students. In this study, the system estimates the head poses and facial expressions of students through the camera and uses the information as criteria for assessing student participation. The estimated result enables the class quality to be categorized into positive, negative, and neutral, thereby allowing teachers to rearrange the class contents. Finally, we evaluate the performance of our system by testing the accuracy of student reaction estimation and our classroom evaluation method.
\end{abstract}

\ccsdesc[300]{Applied computing~Computer-assisted instruction}

\keywords{Online classroom, Classroom evaluation, Human behavior detection, Head pose, Facial expression recognition}


\maketitle

\section{Introduction}
Online education is a new mode of distance education that began in the mid-1990s and emerged with the development of the Internet. Online education has become increasingly highlighted due to its advantages: it transcends the limitations of time and space, and it can mitigate the unequal distribution of educational resources due to geography and other aspects. However, student performance is an independent aspect in any kind of classroom, it is difficult for a teacher to observe each student’s classroom reaction. Moreover, students’ reactions in the classroom are not monolithic; they are a combination of head poses, body movements, and facial expressions. This problem is more prominent in online classrooms. In an online classroom, the teacher must observe the students’ performance from a camera individually to determine the current feedback of the students. However, for an online classroom with 30 students, this problem renders it more difficult for teachers to gauge the students’ reactions in a timely manner and causes the teachers to lose focus on the class content. Moreover, this problem is exacerbated by the multiple reactions of students.

In most current online education systems, teachers can observe 
their students’ class reactions through cameras; however, this 
distracts the teacher from the lesson and requires a significant 
amount of time to observe students’ reactions individually. To 
evaluate the overall classroom quality based on the reaction 
estimation of all students without affecting the attention of 
students and teachers, we developed a new online classroom 
supporting system, that can evaluate student participation based 
on students’ multi-reaction.
 
The remainder of this paper is organized as follows: Section 2 
introduces studies related to student multi-reaction estimation, 
namely, the detection of students’ head poses and facial 
expression. Section 3 describes the process by which an online 
classroom platform is constructed. Section 4 presents the method 
used to estimate students’ multi-reactions in this study, as well as 
the effect of students’ reactions on classroom evaluation. In 
Section 5, a new online classroom evaluation method based on 
students’ multi-reactions is presented. In Section 6, the feasibility 
of the proposed classroom evaluation method based on 
experiments is presented, and the works are discussed. Finally, the 
conclusions are provided in Section 7.

\section{Related Work}

\subsection{Head Pose Estimation}

Humans use their head orientation to convey information during interpersonal interactions, for example, 
a listener nods to a speaker to indicate that he/she understands the information being communicated, 
or a listener pulls his/her head back to indicate avoidance or disapproval. 
Li \cite{Li2017/06} proposed a method that estimates the attention of 30 students in a class using a camera that real-time 
detects their head rotation without recognizing the eyeball pose; 
subsequently, they visualized the three states of student learning. 
Xu \cite{10.1007/978-3-030-37731-1_27} proposed a multiple Euler angle constraint method to create a scoring module to analyze students’ attention 
based on head pose estimation, where the system reported evaluates student attention levels from 0.0 to 1.0.

\begin{figure}[t]
  \centering
  \includegraphics[width=\linewidth]{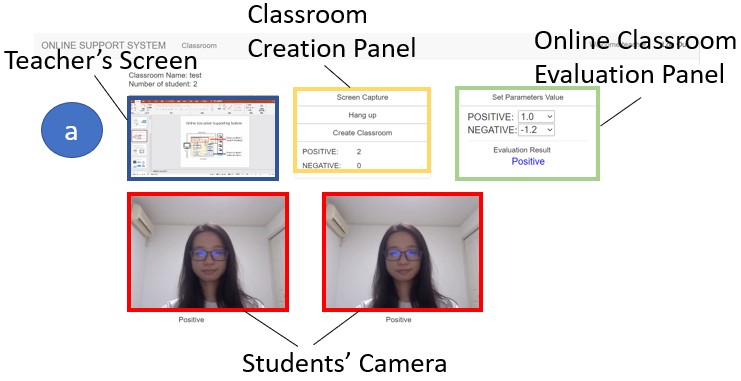}
  \includegraphics[width=\linewidth]{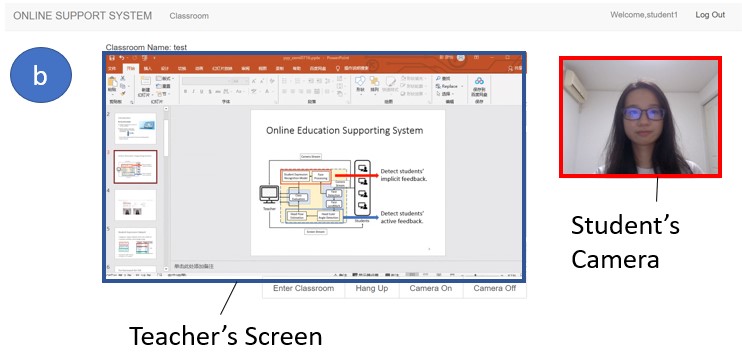}
  \caption{Interface of online classroom system. 
  (a) Teacher page interface containing local stream of teacher’s screen, students’ camera streams transmitted by WebRTC, classroom creation panel, and online classroom panel. 
  (b) Student page interface containing student’s local camera stream and teacher’s shared screen transmitted from teacher’s page by WebRTC.}
\end{figure}

Based on the studies above, it can be concluded that by 
detecting students’ head poses, the current class participation of 
students can be inferred effectively. This study focused on 
detecting each student’s nodding and shaking head poses during 
class via a camera. As only one student is visualized in each 
camera, a multi-target situation does not apply; hence, estimation 
errors are effectively reduced.

\subsection{Facial Expression Recognition}

In addition to head orientation, human expressions convey a 
significant amount of information and emotional states during 
interpersonal communication. Ekman et al. \cite{PMID:5542557} defined six basic 
human expressions and indicated that humans convey the same 
emotional message for basic expressions regardless of culture and 
region. The six basic facial expressions are anger, disgust, fear, 
happiness, sadness, and surprise. In the fields of deep learning and 
computer vision, various facial expression recognition (FER)
systems exist that extract expression information from facial 
representations. With the development of deep learning theory 
and improvement in numerical computing equipment, CNN has
been rapidly developed and widely used in computer vision and 
other fields. Currently, some well-known CNN structures are 
being applied to expression recognition, such as VGGNet \cite{simonyan2015very}, 
which is used to extract image features owing to its brief structure 
and excellent versatility.

\begin{figure}[t]
  \centering
  \includegraphics[width=\linewidth]{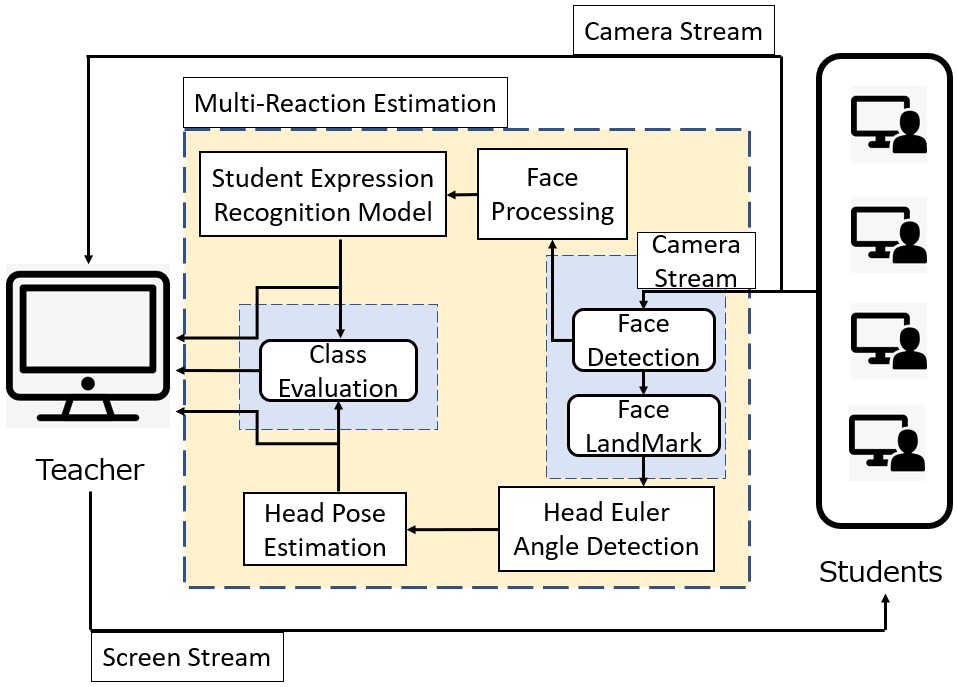}
  \caption{Structure of online classroom system}
\end{figure}

However, in regard to student expressions in the classroom 
environment, the abovementioned criteria for classifying 
expressions are inappropriate. Liang\cite{liang2019intelligent} classified students’ 
expressions into four categories: interest, happiness, confusion, 
and tired according to the classroom environment; and 
automatically recognized facial expressions using a support vector 
machine. However, this study still suffered from a small dataset 
sample and a single scenario.

\section{Online Education Platform}

To provide a platform for teachers and students to conduct online learning activity, we developed a simple online classroom system.

\subsection{Real-time Communication Channel}

Our system is based on WebRTC and sockets for peer-to-peer 
video streaming transmission, which can achieve real-time 
communication through JavaScript provided by browsers. To 
perform educational tasks, the system enables the teacher and 
students to share the teacher’s screen and capture students’ web 
cameras in an online classroom. 

The system allows multiple classrooms to be created based on 
the teachers’ needs, and teachers can create a classroom in the 
classroom creation panel, as shown in Figure 1-a. The different 
classrooms are independent of each other, which allows students 
to enter the classroom by inputting their names.  When the student 
enters the classroom, the student’s socket ID is sent to the teacher 
and recorded by the teacher. In the subsequent student reaction 
estimation, the system distinguishes each student’s detected result 
using this socket ID. After the teacher page receives the socket ID 
from the student page, the teacher and student’s page channels are 
linked. When the channel is connected,
the screen shared by the teacher is displayed on the student 
pages; the interface of the student page is shown in Figure 1-b. On 
the teacher page, the teacher can observe the students’ reactions in
class through a video transmitted from the students’ cameras.
Furthermore, the teacher can verify the students’ responses and 
the overall class evaluation from the online classroom evaluation 
panel.

\subsection{Online Classroom System}

We show the overall structure of our online classroom system in 
Figure 2. First, as described in Section 3.1, the teacher and student 
pages transmit the streams of the teacher-shared page and students’ 
videos through WebRTC and sockets. In addition, the student 
camera video stream is transmitted to the multi-reaction 
estimation module to determine the student’s learning status based 
on the student’s head and facial information. The student’s 
learning status can be determined using one of two approaches: 
head pose estimation and expression recognition.For head pose 
estimation, by detecting the face in the camera video stream, face 
landmarks are obtained, and then the head Euler angle is 
calculated based on the face landmark to determine the head pose. 
For expression recognition, the student’s face region is 
preprocessed after the face region in the camera video stream is 
detected. The expression recognition model is used to detect the 
expression of the processed facial region. Additionally, the current 
student’s listening state (positive, negative, or neutral) is assessed 
based on their head pose and expression.In addition, the 
students’ head poses and expressions are used for classroom 
quality evaluation, and the overall classroom is evaluated by 
evaluating the status of all students.

\section{Multi-Reaction Estimation}

In any type of classroom, both the teacher and students 
significantly affect classroom performance. Students respond to 
the teacher’s course content correspondingly, and the teacher can 
adjust the speed of the course based on the students’ reactions. In 
this study, we classified students’ reactions into positive and 
negative reactions. Students with no feedback regarding the class 
are considered neutral.

\subsection{Student Head Reaction}
Our system assumes that students nod to indicate that they 
understand the teacher’s explanation, and that they shake their 
heads otherwise. The system estimates the students’ head poses 
based on their face landmarks. Therefore, to improve the accuracy 
of head pose estimation, students should position their heads in 
the middle of the video camera area.

To estimate the head pose, the system is required to detect 
faces in video streams transmitted from the students. We used the 
Tiny Face Detector \cite{hu2017finding} as an implement to detect students’ faces in 
the video stream. For each detected face, we can obtain 68 key 
points called face landmarks, and store them in a container of 
points. The front-end sends the students’ face landmarks to the 
server every 100ms. 

Head pose estimation is conducted by obtaining the pose angle 
of the head from the face landmark. In a 3D space, the rotation of 
an object can be represented by three Euler angles: the pitch, yaw,
and roll. To solve the transformation relationship between 2D
facial key points and a 3D face, a 3D face model must be 
developed. In our system, we used a 3D model under normal 
circumstances. To convey 2D information, we used 14 face 
landmarks to create a two-dimensional model. The solvePnP
function provided by OpenCV can calculate the rotation vector 
and translation vector based on two-dimensional facial key points, 
a 3D face model, as well as the camera matrix and
camera distortion. The values of pitch, yaw, and roll can be 
calculated from the rotation vector; subsequently, a simple 
tracking method can be used to estimate the students’ head pose.

Our system estimates the head pose based on the tracking 
method which detects the head between consecutive frames of a 
video stream. Additionally, the head pose is initialized with a 
frontal face to improve the accuracy of head pose estimation 
based on a frontal face. Therefore, we required the students to 
position their heads in the middle of the video camera. The 
general range of motion of the human head are +60 to -60 degrees 
of pitch angle, and +75 to -75 degrees of yaw angle. The research 
of Chen \cite{7406424} demonstrated that the average duration of head 
movements is 850ms. Therefore, we used the following approach 
for head-pose estimation in the ordinary case: (1) Every 100ms, 
we calculate the Euler angle of the students’ head poses. (2) When 
the pitch angle of two adjacent detections is greater than 10 degrees, we 
assume that the student nodded. (3) When the yaw angle of two 
adjacent detections is greater than 12 degrees, we assume that the student 
shook his/her head.

\subsection{Student Expression Reaction}

In Section 4.1, we presented a method to estimate students’ head 
poses based on facial landmarks; this method can effectively help 
teachers assess the current students’ understanding of the course 
content. However, 
the estimation of head poses requires students’ active feedback; 
this implies that if students do not actively provide feedback 
regarding the classroom content, then our system is not able to 
support the teachers’ evaluation of the classroom. Therefore, we 
propose the detection of students’ implicit feedback regarding 
classroom content based on student expressions.

As mentioned in Section 2.2, the classification of the six basic 
expressions (anger, disgust, fear, happiness, sadness, and surprise) 
cannot be applied simply to the educational 
environment. In this study, we primarily categorized classroom 
expressions into happiness, focused, confused, disgust, and tired 
based on the students’ emotions in the classroom. Finally, we 
defined the students who had no expression changes as neutral.

Most existing expression recognition datasets are based on six 
basic expressions for classification. To create a suitable dataset for 
recognizing classroom expressions, we obtained expressions in 
the following two categories: For part A, which is associated with 
basic expressions of happy, disgust, and neutral, we selected 
expression samples from the existing datasets; for part B, which is 
not associated with basic expressions of focused, confused, and 
tired, we obtained samples from Google images.

In this study, we selected happiness, disgust, and neutral 
samples from three datasets: JAFFE\cite{lyons1998coding}, CK+\cite{5543262}, and SFEW 
2.0\cite{10.1145/2818346.2829994}. Among them, JAFFE and CK+ are laboratory-controlled 
samples, whereas SFEW 2.0 is intercepted from the actors’ 
expressions in the movie clips. Because CK+ is a sequence dataset, 
we extracted the last frame with peak formation and the first three 
frames (for neutral face) of each sequence. For samples in part B, 
we used the following approach to obtain the expression samples: 
First, we collected images of keywords through the Selenium API;
next, we filtered non-face images using the Tiny Face Detector. 
Finally, we manually selected images that did not match the 
keywords (confusion expression, focused expression sleepy, and
sleepy expression) or education environment. For example, we
removed some exaggerated expressions and some images with too 
many facial obscurations.

The dataset of student expressions in the classroom that we 
collected is shown in Table 1. After organizing the dataset, we 
cropped the face region and performed grayscale processing; 
finally, the sample image was resized to 48 × 48. Figure 3 shows 
an example of each expression after pre-processing. Similar to the 
evaluation benchmark of student reactions presented in Section 
4.1, we reclassified the student expression dataset as positive, 
negative, or neutral. Among them, happiness and focused were 
positive reactions; disgust, confused, and tired were negative 
reactions.

\begin{figure}[t]
  \centering
  \includegraphics[width=0.6\linewidth]{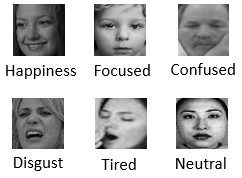}
  \caption{Examples of datasets used for student expression 
  recognition. In these examples, focused, confused, and tired 
  samples from Google image. Happiness and disgust samples 
  from SFEW 2.0. Neutral sample from JAFFE.}
\end{figure}

\begin{figure}[t]
  \centering
  \includegraphics[width=\linewidth]{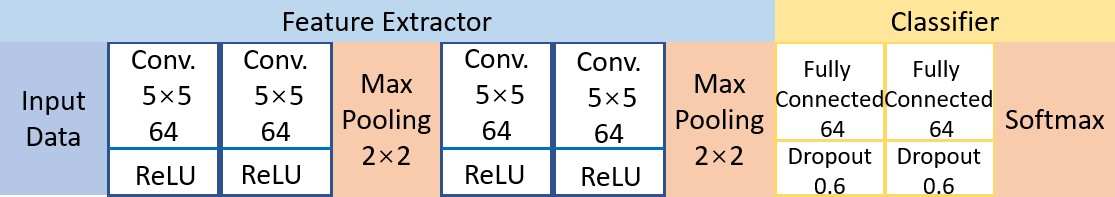}
  \caption{The architecture of CNN model}
\end{figure}

The CNN structure that we constructed based on ours frame \cite{kuo2018compact},
which contains four convolution layers with two 
additional fully-connected layers at the end of the network and a 
ReLU were used for each convolution layer as an activation 
function. The Figure 4 shows the architecture of our CNN model.

\section{Online Classroom Evaluation}

In our system, class evaluation is based on the students’ reaction 
estimation. After the system estimates the students’ head poses 
and expression reactions, the system collects the head poses and 
the expression reactions of all students in the classroom to
evaluate the current classroom participation. During class, we 
assume that a) the class content is related before and after, and b) 
some students are not attentive. For those cases, we assume that 
when the student does not understand the previous content of the 
course, it is more difficult for him/her to understand the current
explanations of the teacher as compared with other students. For 
our evaluation method, we assume that students with good 
performance participation are more important in the classroom. 
Therefore, we increase the effect of positive reactions on the class 
evaluation when considering the class evaluation method. To 
reduce interference in the two abovementioned cases, a weighted 
method is used in the system to calculate the overall listening of 
the students.

\begin{table}
  \caption{Our dataset of student expression for classroom 
  expression recognition, it contains 1,089 training set and 281 
  testing set. }
  \label{tab:freq}
  \begin{tabular}{l*{5}{c}}
    \toprule
    Category& Samples & Train & Test & Total\\
    \midrule
    Happiness & 216 & 170 & 46 & \multirow{6}*{1370}\\
    Focused & 85 & 68 & 17& \\
    Confused & 134 & 106 & 28 & \\
    Disgust & 130 & 101 & 29 & \\
    Tired & 96 & 76 & 20 & \\
    Neutral & 709 & 568 & 141 & \\
  \bottomrule
\end{tabular}
\end{table}

As mentioned in Section 4, we regard students’ nodding heads, 
happiness, and focused as positive reactions; and shaking head, 
disgust, confused and tired as negative reactions. We evaluate the 
quality of the class $r_{class}$ using the following method:
\begin{equation}
  r_{class}= \sum_{i=1}^{n} w_{i}r_{i}
\end{equation}

Where $r_{i}$ and $w_{i}$ are the reaction and weight of student $s_{i}$,
respectively. $r_{class}$ shows the value of a class reaction considering 
$r_{1},r_{2},r_{3},...,r_{n}$.

We consider that when the majority of students in the 
classroom are positive, the current classroom evaluation result is 
positive, on the contrary, we consider the classroom as negative if 
negative students are the majority of the classroom. Therefore, 
when $r_{class}$ is greater than or equal to 0.2, the current classroom 
level is considered positive. When the $r_{class}$ is less than -0.15, the 
current classroom level is considered as negative. When the $r_{class}$
is -0.15 to 0.2, the current classroom level is neutral. For cases 
mentioned above, the method evaluates the current class reaction 
by considering both students’ past performance and current 
reaction. This method can reduce the effects of students who have 
not participated in the past on the current classroom evaluation. 
The classroom evaluation method is suitable for small classes 
comprising 30 students or less.

Here, we show an example of POSITIVE result on the teacher 
page in Figure 5. To define the value of $r_{i}$, we set 1 for positive 
reactions, -1.2 for negative reactions, and 0 for neutral. We 
defined the positive and the negative reactions with unequal 
values because in a classroom, teachers should conduct classroom 
activities that enable students to understand the content of the 
lesson; therefore, we assigned a higher negative value to negative 
reactions.

For weight of each student, we define the value of $w_{i}$ as follows:
\begin{equation}
  w_{i}=\frac{f_{i}}{\sum_{j=1}^{n}f_{j}}
\end{equation}

This weight calculation algorithm reduces the weight 
coefficient $f_{i}$ proportion of student $s_{i}$ as the number of students 
increases. When the number of students in the classroom is
sufficiently large, the weight coefficient $f_{i}$ proportion of student
$s_{i}$ approaches 0. Therefore, we specify less than 30 students in the 
classroom when using this method to assess classroom level.

Then, count the number of negative reactions $n_{i}$ and the 
number of positive reactions $p_{i}$ of the student $s_{i}$ from the 
beginning to the positive moment. In addition, we considered the 
effect of the number of student reactions on the classroom 
evaluation function. Therefore, we used a logarithmic function to 
reduce the effect of the students’ reaction times.

The weight coefficient $f_{i}$ of student $s_{i}$ is expressed as
\begin{equation}
  f_{i}=\log \left ( 2p_{i}+n_{i} \right )
\end{equation}

\begin{figure}[t]
  \centering
  \includegraphics[width=0.8\linewidth]{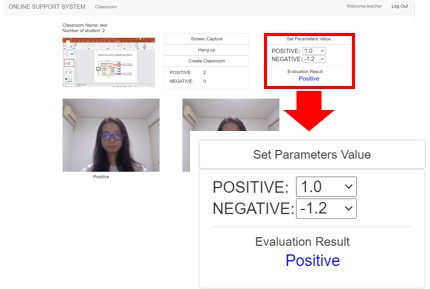}
  \caption{An example of POSITIVE Result on the teacher page}
\end{figure}

We use the logarithmic function to reduce the effects of students’ 
past reaction times on the weight coefficient, where $p_{i}$ is twice of $n_{i}$. 
This is because we assume that in past reactions, students with 
more positive reactions can better understand the content taught 
by the teacher, and hence it is relatively easy for them to 
understand the current class content. As such, the coefficient of 
positive reaction $p$ has a higher weighting than the negative reaction $n$.

\section{Evaluation and Discussion}

To assess the performance of our online classroom support system, 
we firstly evaluated our head pose estimation and student 
expression recognition model, and the test results were used to 
verify the class evaluation method proposed in Section 5.

\subsection{Evaluation of Reaction Estimation}

We evaluated our head pose estimation experimentally. We 
performed 30 discontinuous head nodding and 30 discontinuous 
head shaking on the student’s page, and the experimental results 
obtained on the teacher page are shown in Table 2. In the 30-
nodding-head tests, 25 were correctly predicted as nodding, 3
were predicted as shaking, and 2 were neutral; the accuracy of the 
nodding test was 0.83. Meanwhile, in the 30-shaking-head tests, 
24 were correctly predicted as shaking, 5 were predicted as 
nodding, and 1 was neutral; the accuracy of the shaking test was 0.80.

Next, we describe our experiment using the proposed 
expression recognition model. The dataset mentioned in Section 
4.2 was categorized into a training set and a testing set, and they
contained 1,089 and 281 samples, respectively. We trained our 
expression recognition model using the dataset obtained, and the
experimental results are shown in Table 3. This experiment 
involved 84 negative samples, 49 positive samples, and 148 
neutral samples. For the negative, positive, and neutral samples, 
the final accuracies of the model were 0.71, 0.79, and 0.89, respectively.

\begin{table}
  \caption{Experimental result of head pose estimation}
  \label{tab:freq}
  \begin{tabular}{c|cc}
    \hline
    \diagbox{Result}{Action} & Nodding Head & Shaking Head \\
    \hline
    Nodding Head & 25 & 5\\
    Shaking Head & 3 & 24\\
    Neutral & 2 & 1\\
    \hline
    Accuracy& 0.83 & 0.80\\
    \hline
\end{tabular}
\end{table}

\begin{table}
  \caption{Experimental result of expression recognition model}
  \label{tab:freq}
  \begin{tabular}{c|ccc}
    \hline
    \diagbox{Result}{Input} & Negative & Positive & Neutral \\
    \hline
    Negative & 60 & 6 & 11\\
    Positive & 19 & 38 & 6\\
    Neutral & 5 & 5 & 131\\
    \hline
    Accuracy & 0.71 & 0.79 & 0.89\\
    \hline
\end{tabular}
\end{table}


\subsection{Classroom Evaluation}

Based on the experimental result of student reaction estimation, 
we used the method mentioned previously to evaluate the class as 
a whole. Because the number of students in the classroom, such as
10, 22, or 30 students, as well as the reaction of students can 
affect the classroom evaluation results, we assumed a classroom 
of 30 students, and that the number of students’ previous state 
reaction was between 6 and 12 times. Among them, 16 students 
had more positive than negative reactions, 10 students had more 
negative than positive reactions, and four students had the equal 
number of positive and negative reactions. Next, considering the 
students’ current reactions, we changed the classroom evaluation 
$r_{class}$ by adjusting the students’ current reactions. To evaluate the 
classroom, we used the test results presented in Section 6.1, and
assumed that the probability of students performing head pose 
reaction and expression reaction was equal.

By randomly adjusting each student’s current reaction state 
(positive, negative, and neutral), we tested the true $r_{class}$ from -1.2 
to 1 for each case with a step size of approximately 0.1, and 
each case was tested 1,000 times to obtain the average $r_{class}$ and 
the classroom level prediction; the results are shown in Figure 6. 
As an example, when all students were positive, the true $r_{class}$ is 1. 
However, as presented in Section 6.1, the system’s estimation of 
student reactions differ from the real student reaction, and our 
classification of the overall classroom level, the $r_{class}$ of these
1,000 tests were all positive; therefore, we assumed that the 
accuracy of the proposed method was 100$\%$ in evaluating the 
toward the zero as compared with the true $r_{class}$ value. When the 
true $r_{class}$ was far from zero, the test $r_{class}$ deviated significantly 
from the true $r_{class}$, and the maximum deviation value was 0.509. 
When the true $r_{class}$ was approximate to zero, the test $r_{class}$ did 
not deviate significantly from the true $r_{class}$, and the minimum 
deviation value was 0.007. In addition, the accuracy of our 
proposed method for evaluating the classroom was high when the 
true value was far from the discriminant value (0.2 and -0.15), and 
the highest accuracy was 1. By contrast, when the true value was 
approximately the discriminant value, the accuracy of the method 
for evaluating the classroom was low, with the lowest accuracy of 
0.267. The average accuracy of the proposed method was 0.852.
Considering another situation where the weight $w_{i}$ of students is 
not affected by previous behavior, that is, the weight between 
students is equal, we got the following results, the average 
accuracy of no weight situation is 0.842, it is lower than the 
method we proposed.

\begin{figure}[t]
  \centering
  \includegraphics[width=0.8\linewidth]{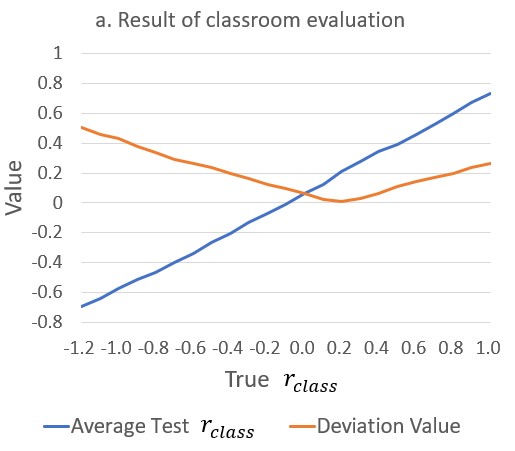}
  \includegraphics[width=0.8\linewidth]{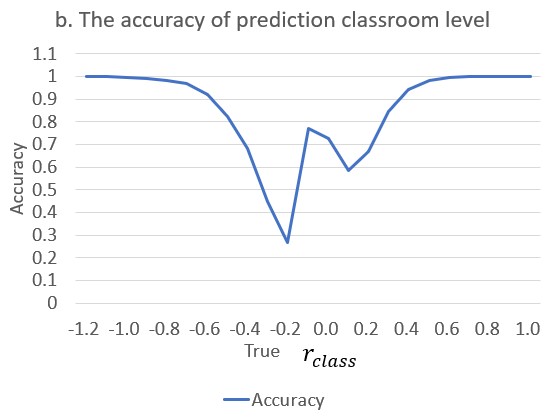}
  \caption{(a) Result of classroom evaluation. (b) Accuracy of classroom level prediction.}
\end{figure}

\subsection{Discussion}
For student reaction estimation, the estimation accuracy of our 
system is to be improved.
For head pose estimation, the system requires students to position 
their heads in the middle of the camera, and the system cannot 
easily detect head rotations that are extremely subtle. Regarding 
the laboratory-controlled facial expression datasets we used for 
student expression recognition, some of the facial expressions 
were exaggerated and did not match the students’ facial 
expressions in the classroom environment. In addition, the 
methods for estimating head poses and facial expressions have 
high requirements for the background and illumination of the 
image captured by the camera. In future studies, the 
abovementioned aspects should be improved, and evaluation 
experiments should be strengthened by using real videos of 
students in a class.

\section{Conclusion}
A new method for assessing student participation through head 
pose estimation and expression recognition was proposed herein; 
in this method, the current classroom is evaluated based on 
students’ reactions. This study focused on the classroom 
evaluation method, and experimental results showed that the 
proposed method can effectively evaluate the overall listening of 
students with an accuracy of 85.2$\%$, even though the reaction 
estimation for a few students was inaccurate. Therefore, the 
system can support teachers in providing better teaching activities 
and evaluating small online classes.

\begin{acks}
This work was supported in part by JSPS KAKENHI Grant Number JP19K12266.
\end{acks}

\bibliographystyle{ACM-Reference-Format}
\bibliography{sample-base}

\end{document}